\documentclass[fleqn,usenatbib]{mnras}

\usepackage{newtxtext,newtxmath}

\usepackage[T1]{fontenc}

\DeclareRobustCommand{\VAN}[3]{#2}
\let\VANthebibliography\thebibliography
\def\thebibliography{\DeclareRobustCommand{\VAN}[3]{##3}\VANthebibliography}

\usepackage{graphicx}	
\usepackage{amsmath}	
\usepackage{soul}

\newcommand{\src}{4U~1543--475}
\newcommand{\nustar}{\textit{NuSTAR}}
\newcommand{\nicer}{\textit{NICER}}
\newcommand{\hxmt}{\textit{Insight}-HXMT}
\newcommand{\meerkat}{\textit{MeerKAT}}
\newcommand{\maxi}{\textit{MAXI}}
\newcommand{\swift}{\textit{Swift}}

\usepackage[normalem]{ulem}

\title[Radio flares in \src]{Radio flares and X-ray hardening embedded in the long soft state of \src}
\author[Zhang et al.]{Zuobin Zhang,$^{1}$\thanks{E-mail: zuobin.zhang@physics.ox.ac.uk}
Rob Fender,$^{1,2}$
Jiachen Jiang,$^{3}$
Payaswini Saikia,$^{4}$
David M. Russell,$^{5}$ 
Andrew Hughes,$^{1}$ \newauthor
Honghui Liu,$^{6}$  
Francesco Carotenuto,$^{7}$
James F. Steiner,$^{8}$ 
Fraser J. Cowie,$^{1}$ 
John A. Tomsick,$^{9}$ \newauthor
Cosimo Bambi,$^{10,11}$ 
Yimin Huang,$^{10}$
Xian Zhang,$^{12}$
Wenfei Yu,$^{12}$
Yuexin Zhang,$^{8,13}$ 
and Rittick Roy,$^{14}$
\\
$^{1}$Astrophysics, Department of Physics, University of Oxford, Keble Road, Oxford OX1 3RH, UK\\
$^{2}$Department of Astronomy, University of Cape Town, Private Bag X3, Rondebosch 7701, South Africa\\
$^{3}$Department of Physics, University of Warwick, Gibbet Hill Road, Coventry CV4 7AL, UK\\
$^{4}$Department of Astronomy, Yale University, PO Box 208101, New Haven, CT 06520-8101, USA\\
$^{5}$Center for Astrophysics and Space Science (CASS), New York University Abu Dhabi, PO Box 129188, Abu Dhabi, UAE\\
$^{6}$Institut f\"ur Astronomie und Astrophysik, Eberhard-Karls Universit\"at T\"ubingen, D-72076 T\"ubingen, Germany\\
$^{7}$INAF-Osservatorio Astronomico di Roma, Via Frascati 33, I-00078, Monte Porzio Catone (RM), Italy\\
$^{8}$Center for Astrophysics, Harvard \& Smithsonian, 60 Garden St, Cambridge, MA 02138, USA\\
$^{9}$Space Sciences Laboratory, 7 Gauss Way, University of California, Berkeley, CA 94720-7450, USA\\
$^{10}$Center for Astronomy and Astrophysics, Department of Physics, Fudan University, Shanghai 200438, China\\
$^{11}$School of Natural Sciences and Humanities, New Uzbekistan University, Tashkent 100000, Uzbekistan\\
$^{12}$Shanghai Astronomical Observatory, Chinese Academy of Sciences, Shanghai 200030, China\\
$^{13}$Kapteyn Astronomical Institute, University of Groningen, P.O. BOX 800, 9700 AV Groningen, The Netherlands\\
$^{14}$Anton Pannekoek Institute for Astronomy, University of Amsterdam, Science Park 904, NL-1098 XH Amsterdam, the Netherlands}

\date{Accepted XXX. Received YYY; in original form ZZZ}

\pubyear{\the\year{}}

\begin{document}
\label{firstpage}
\pagerange{\pageref{firstpage}--\pageref{lastpage}}
\maketitle

\begin{abstract}
We present a comprehensive multi-wavelength study of the black hole X-ray binary \src\ during its 2021 outburst, focusing on radio flaring episodes that are commonly interpreted as signatures of episodic jet production and are embedded within states when the X-ray emission was dominated by an accretion disk component. The radio monitoring reveals at least two discrete flares that coincide with periods of enhanced Comptonized X-ray emission. Broadband spectral modelling shows a significant decrease in the reflection-to-disk flux ratio (by a factor of $\sim3-4$) during these episodes, consistent with a temporary change in the geometry of the inner accretion flow, although the data do not allow the causal sequence to be firmly established. Optical photometry exhibits variability that broadly tracks the reflection fraction, consistent with changes in the illuminating component. The accompanying spectral hardening indicates that the radio flares were associated with short-lived excursions toward a "harder" state, departing from the soft state. X-ray timing analysis suggests that the radio flares may be associated with changes in the fractional rms variability; however, no consistent or unified pattern can be firmly established across different events. These results provide a multi-wavelength observational example of radio flaring activity in a black hole binary and highlight the complex interplay between accretion flow geometry, coronal emission, and jet-related phenomena.
\end{abstract}




\section{Introduction} \label{sec:intro}

Relativistic jets are among the most energetic and fascinating phenomena produced in accreting black holes, appearing across an enormous range of black hole masses—from active galactic nuclei (AGN) to stellar-mass black hole X-ray binaries (XRBs). In both systems, the coupling between the accretion flow and jet outflow reveals striking similarities once the characteristic timescales are scaled by black hole mass \citep{Merloni2003, Falcke2004, Fender2007, Saikia2015, McHardy2006}. Despite decades of study, the physical connection between the accretion flow and jet production remains an open question.

In black hole XRBs, distinct X-ray spectral states correspond to different accretion regimes and jet properties \citep{Kalemci2022, Remillard2006, Fender2004}. During the hard state, the X-ray spectrum is dominated by Comptonized coronal emission and a compact, steady jet is present \citep{Sunyaev1979, Fender2004}. As the system transitions toward the soft state, the corona emission weakens, the accretion disk may extend inward, and discrete, relativistic ejections are often observed near the transition. The subsequent soft state is typically characterized by a disk-dominated X-ray spectrum and weak variability, with the compact jet quenched \citep{Shakura1973, Fender1999, Coriat2011, Russell2011, RussellT2019}. Previous and more recent studies suggest that some sources can exhibit jet activity in their disk-dominated state \citep[e.g.][]{Brocksopp2013, Williams2022}, as an alternative outflow form of disk winds \citep{Zhang2026}. High-cadence multi-band observations of systems such as MAXI~J1820+070 have revealed that transient jet-launching episodes coincide with dramatic changes in X-ray timing properties—specifically, the replacement of strong type-C quasi-periodic oscillations (QPOs) by type-B QPOs and a sharp drop in fractional root mean square (rms) variability \citep{Homan2020}. The connection between drops in X-ray variability and jet ejections has been suggested in some sources \citep{Fender2009, Miller-Jones2012, Carotenuto2025}.


Observations of radio-loud AGN such as 3C~120 and 3C~111 have revealed associations between powerful jet ejections and reductions in X-ray luminosity \citep{Marscher2002, Chatterjee2009, Chatterjee2011}. In light of these observations, a “jet cycle” has been proposed, involving a geometrically thin disk extending to the ISCO, the onset of instabilities in the inner disk, the launch of a transient relativistic outflow, and the refilling of the innermost accretion region \citep{Lohfink2013, Fedorova2023}, although the precise sequence of physical events in these systems is not unambiguously established. These AGN results were originally interpreted through analogies with black hole X-ray binaries, where similar phenomenology had been identified earlier \citep[e.g.][]{Pooley1997, Fender1997, Rodriguez2008, Mendez2022}. Understanding radio events in XRBs is therefore crucial, not only for constraining how jet ejections are triggered in these systems, but also for providing key insight into the processes that govern powerful jet production in AGN. Nonetheless, the detailed physical connection between accretion flow and jet launching remains uncertain, particularly for flares occurring during predominantly soft accretion states, where the jet mechanism may differ from that associated with canonical hard-to-soft transitions.

The black hole XRB \src\ offers an excellent laboratory for probing into this issue. It is a well-studied, high-luminosity transient system ($M_{\rm BH}$=$9.4$~$M_{\bigodot}$, $D=5-7.5$~kpc) \citep{Orosz1998, Russell2006, Zhang2025_2} that has undergone several outbursts since its discovery, exhibiting both strong thermal disk emission and episodic radio and infrared jet flares \citep{Russell2020, Zhang2025}. Its relatively low inclination ($\sim 20^\circ$; \citealt{Orosz2003, Zhang2025_2}) allows a clear view of the inner accretion flow, while its rich observational coverage in X-ray and radio bands enables detailed studies of accretion–ejection coupling. \citet{Zhang2025_2} reported two spatially resolved, high Lorentz factor ($>$4.6 at launch) ejections from the system, demonstrating that X-ray binaries can launch jets with velocities comparable to those observed in AGN, as suggested by the statistical study \citep{Lilje2025}.

In this work, we present a comprehensive analysis of the episodic radio flares during the long-lived disk-dominated state of \src\ using \nustar, \nicer, and \hxmt\ observations combined with the \meerkat\ observations reported by \citealt{Zhang2025_2}. Our results suggest a broad disk–jet coupling scenario associated with episodic radio flares.


The paper is organized as follows. In Sec.~\ref{sec:data}, we present the collection of observational data and the data reduction. The spectral fitting results are reported in Sec.~\ref{sec:analysis}. We discuss the results and report our conclusions in Sec.~\ref{sec:discussion} and Sec.~\ref{sec:conclusion}, respectively.


\begin{figure*}
    \centering
    \includegraphics[width=0.98\linewidth]{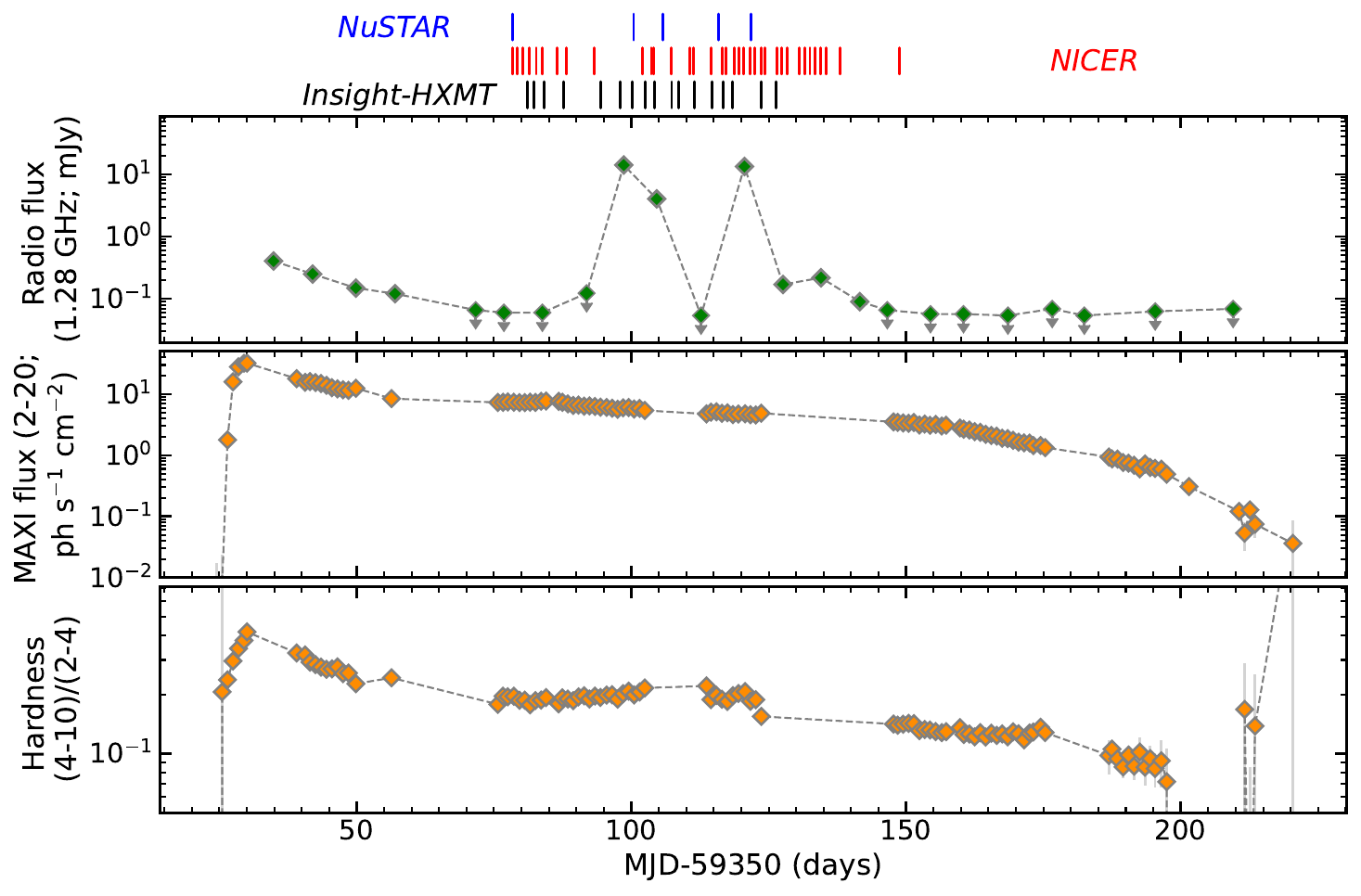} \\
    \caption{multi-wavelength light curves of \src\ during the 2021--2022 outburst. Top panel: 1.28~GHz radio flux density measured with \meerkat. Error bars represent the 1~$\sigma$ uncertainties, which are smaller than the symbol size and therefore not visible. Non-detections are shown as 3~$\sigma$ upper limits with downward arrows. Middle panel: \maxi\ 2--20~keV X-ray flux with 1~$\sigma$ confidence interval error bars. Bottom panel: \maxi\ hardness ratio (4--10~keV / 2--4~keV). The vertical lines at the top of the figure mark the epochs of \nustar\ (blue), \nicer\ (red), and \hxmt\ (black) observations. As reported by \citet{Jin2024} and \citet{Yorgancioglu2023}, the source underwent a transition from a super-Eddington state to a sub-Eddington state around MJD=59405. We restrict our analysis to observations obtained after this state transition, i.e. during the sub-Eddington phase, which includes the radio flaring interval of interest.}
    \label{maxi_lc}
\end{figure*}

\section{Observation and data reduction} \label{sec:data}

\src\ entered a new outburst phase in June 2021, as reported by \maxi\ and \swift\ \citep{Negoro2021_1}. A hard-to-soft spectral state transition began shortly after the initial detection, resembling its previous outburst \citep{Park2004}. The source reached a peak flux of about 9~Crab, close to its Eddington luminosity, making it the brightest black hole transient observed in over a decade \citep{Negoro2021_2}.

\subsection{Radio observations}

Following the confirmation of the outburst, \meerkat\ initiated regular radio monitoring of the source at a cadence of approximately one week as part of the ThunderKAT Large Survey Programme \citep{Fender2016}. The resulting 1.28~GHz (L-band) flux densities were published in \citet{Zhang2025}; see also the top panel of Figure~\ref{maxi_lc} for the corresponding light curve of the core radio emission covering the main outburst phase. We note that the radio emission discussed here originates from jet activity at the source core position. Starting from MJD~59462, an additional radio component was detected at a projected separation of $\sim9$~arcsec from the core. Follow-up observations indicate that this component is consistent with a transient jet launched at an earlier time (assumed to be MJD~59376; \citealt{Zhang2025_2}). Given the angular resolution of \meerkat\ at 1.28~GHz (typically $\sim5$ arcsec), we can safely rule out any significant contribution of this ejecta to the radio emission measured at the core position.

\meerkat\ detected enhanced radio activity between MJD~59448 and 59491, which we focus on in this work. All detections show negative spectral indices between -1.5 and -0.5 \citep{Zhang2025}, consistent with optically thin emission from transient jets with discrete ejections \citep[e.g.][]{Fender1999_2, Russell2019}. The peaks of two flare events are separated by approximately 22 days. Assuming a proper motion of 180~mas/day, typical for jet ejecta in \src\ \citep{Zhang2025}, the ejecta associated with the first flare, if there was one, would be expected to reach an angular separation of $\sim4$ arcsec from the core after 22 days. This scale is comparable to the angular resolution of \meerkat\ at 1.28~GHz. In such a case, the ejecta may appear as marginally extended or elongated emission rather than as clearly resolved components. However, no such extended structure is observed in the \meerkat\ image corresponding to the second flare. This suggests that the second flare is unlikely to be associated with the same ejecta, and instead points to an independent episode of activity originating from the core. In addition, the relatively sparse cadence of the radio observations (approximately one observation per week) may result in short-lived flaring events being missed \citep{Fender2023}. These considerations suggest that there are at least two distinct jet ejection events from the core during this period.


In parallel, \maxi\ provided daily X-ray monitoring. Figure~\ref{maxi_lc} presents the 2–20~keV daily-averaged X-ray light curve and the corresponding hardness ratio (4–10~keV / 2–4~keV), offering a broad view of the source’s flux evolution during the jet ejection periods. The \maxi\ light curves were obtained from the on-demand data system \citep{Matsuoka2009}. As indicated in the same figure, these intervals of increased radio activity were also observed with long exposures by \nustar, \nicer, and \hxmt, which constitute the main data sets used in this study.

\subsection{\nustar\ data reduction}

A total of 11 \nustar\ observations are available for this source. Following the state classification of \citet{Jin2024} and \citet{Yorgancioglu2023}, which showed that the source transitioned from a super-Eddington state to a sub-Eddington state around MJD=59405, we select the five observations (ObsIDs: 907023260-04, -06, -08, -10, and -12) obtained after this transition, as they correspond to the sub-Eddington phase that encompass the radio flaring interval of interest. The observational data are reduced following the standard pipeline using the \nustar\ Data Analysis Software (NUSTARDAS v2.1.1) and the latest calibration files from CALDB v20220301. Calibrated and cleaned event files are produced with the task \texttt{nupipeline}. We set \texttt{saamode=STRICT} and \texttt{tentacle=YES} to exclude the passages through the South Atlantic Anomaly (SAA). For this extremely bright source, we apply \texttt{statusexpr="STATUS==b0000xxx00xxxx000"} as recommended \footnote{https://heasarc.gsfc.nasa.gov/docs/nustar/analysis/}. Source events are extracted from a circular region with a radius of $150''$ centered on the brightest pixel, while background events are taken from a nearby source-free circular region of the same size. The source and background spectra, light curves, and response files are generated separately for FPMA and FPMB using the task \texttt{nuproducts}.

\subsection{\nicer\ data reduction}

\nicer\ began intensive, high-cadence monitoring of the source on 2021 June 12, one day after the outburst discovery, and continued throughout the entire outburst. We include 40 \nicer\ observations in this study (ObsID: 4202230140--4202230188, except ones with an exposure time less than 100~s). The data are processed using the \nicer\ Data Analysis Software (NICERDAS v11) and calibration database (CALDB v20221001). We conduct the standard \nicer\ reduction routine \texttt{nicerl2} to reduce the data with the default filtering criteria. Events flagged as "undershoot" or "overshoot" (\texttt{EVENT\_FLAGS=bxxxx00}) are excluded. Additionally, we remove the data of detectors \# 14 and \# 34 because of electronic noise. The energy spectra of the background is extracted with the \texttt{nibackgen3C50} tool \citep{Remillard2022}. We create the source products, Redistribution Matrix File (RMF), and Ancillary Response File (ARF) using the tasks \texttt{nicerrl3}, respectively.

\subsection{\hxmt\ data reduction}

\hxmt\ covers a broad energy range of 1--150~keV with its low-energy (LE), medium-energy (ME), and high-energy (HE) detectors. It has observed the 2021 outburst of \src\ during 41 different epochs, among which 18 epochs are within the period we are interested in (ObsIDs: P0304026024-P0304026041). We process the data using the \hxmt\ Data Analysis Software (HXMTDAS v2.06) and the calibration database (CALDB v2.07), following the official user guide \footnote{http://hxmten.ihep.ac.cn}. Background estimation is performed with the dedicated tools \texttt{mebkgmap} and \texttt{lebkgmap}, respectively \citep{Liao2020, Liao2020_2}. We screen the data using the standard criteria as suggested by \hxmt\ manual. Due to the low signal-to-noise ratio of the HE (35--150~keV) data, the HE data are not included for spectral analysis.



\section{Spectral analysis and results} \label{sec:analysis}

The \nustar\ and \nicer\ spectra are grouped with a minimum of 50 counts per bin using the task \texttt{ftgrouppha}, while \hxmt\ are grouped to ensure a minimum count of 20 per bin. We analyze the spectra using the X-ray spectral fitting package XSPEC 12.13.1 \citep{Arnaud1996}. $\chi^2$ statistics are employed and all parameter uncertainties are estimated at 90\% confidence level, corresponding to $\Delta\chi^2=2.71$. We utilize the Wilm set of abundances \citep{Wilms2000} and Vern photoelectric cross sections \citep{Verner1996} in all fits. 

\subsection{Broadband spectral model}

The 2021 outburst of the black hole X-ray binary \src\ has provided an unprecedented opportunity to study accretion physics across a broad energy range. Multiple studies using \nicer, \nustar, and \hxmt\ data have performed detailed broadband spectral fitting to constrain the geometry, ionization state, and physical conditions of the accretion flow \citep[e.g.][]{Yang2024, Jin2024, Zhao2024}. In this work, we aim to interpret all observations within a unified modeling framework, incorporating data obtained from different instruments. To establish such a consistent model framework, we analyze the broad-band X-ray spectra obtained from the simultaneous or quasi-simultaneous \nustar\ and \nicer\ observations ($\Delta$MJD$<2$). During this period (sub-Eddington soft state), a total of five \nustar\ observations were performed (labeled as Epoch~1 to Epoch~5), each accompanied by a simultaneous or quasi-simultaneous \nicer\ observation. We use 0.5--10~keV data for \nicer\ spectra and 3--79~keV data for \nustar\ spectra.

We initially fit the \nustar\ spectra with a model combination of disk blackbody component (\texttt{diskbb}; \citealt{Makishima1986}) and thermal Comptonization component (\texttt{nthcomp}; \citealt{Zdziarski1996}), \texttt{const $\times$ tbnew\_feo $\times$ (diskbb+nthcomp)} in XSPEC language. The multiplicative constant model (\texttt{const}) is included to account for the normalization discrepancy between diﬀerent instruments, while absorption model \texttt{tbnew\_feo} is included to describe the Galactic absorption. The seed photon temperature ($kT_{\rm bb}$) in \texttt{nthcomp} is linked to the inner disk temperature ($kT_{\rm in}$) of \texttt{diskbb}. To reveal any potential reflection features, we temporarily exclude the 4--9~keV and 10--20~keV energy ranges from the fit, while fixing the interstellar absorption column density at $0.4\times10^{22}$~cm$^{-2}$ following \citet{Park2004} and \citet{Morningstar2014}. The best-fitting continuum model was then extrapolated into the excluded energy ranges.

\begin{figure}
    \centering
    \includegraphics[width=1.0\linewidth]{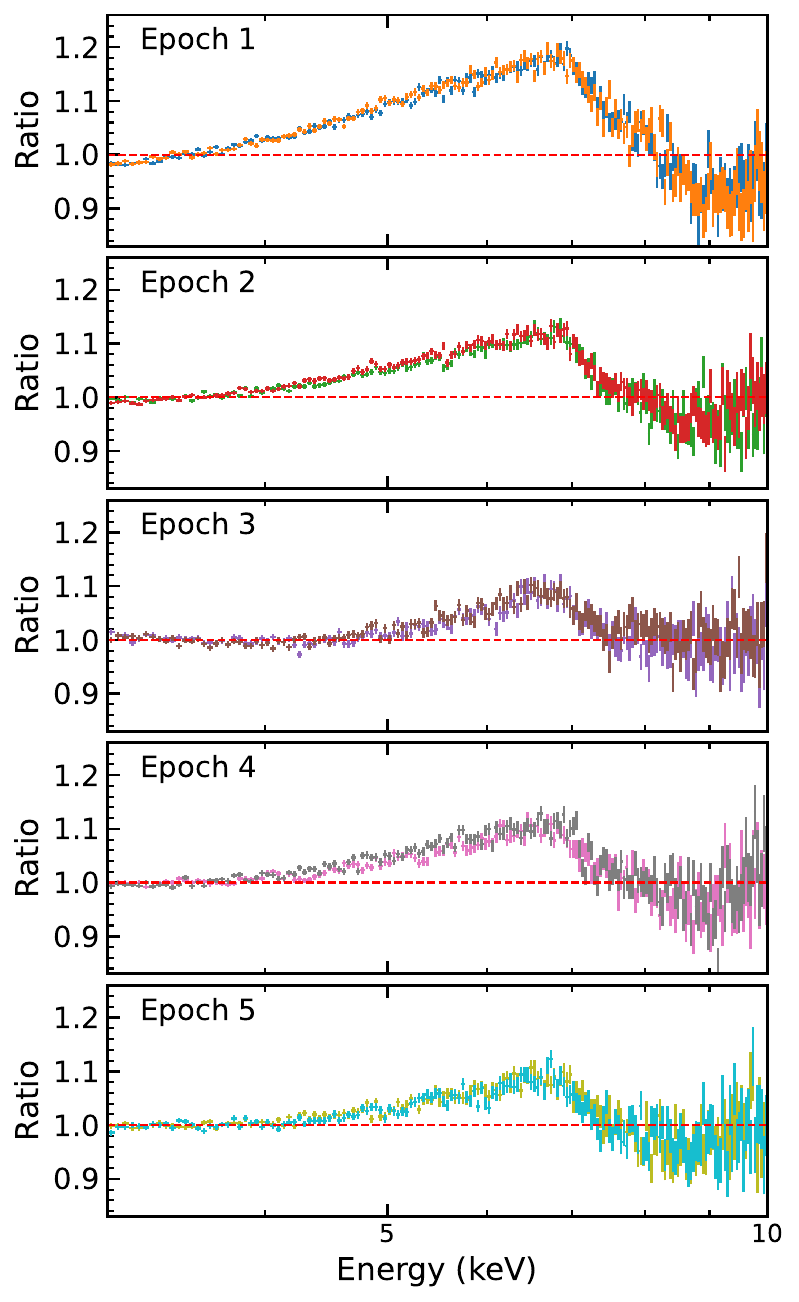} \\
    \caption{Residuals of the \nustar\ spectra obtained by extending the thermal + Comptonization model to the iron-line energy range. The two colors represent data from FPMA and FPMB, respectively. A broad emission feature peaking at $\sim$6.7 keV is evident in all epochs, showing similar profiles across observations but with noticeable differences in width and strength. These variations likely reflect structural changes in the innermost accretion disk.}
    \label{nustar_residuals}
\end{figure}

Figure~\ref{nustar_residuals} shows the residuals obtained when extrapolating the thermal + Comptonization model to energy range covering the whole iron-line. The plot clearly reveals a broad emission feature peaking at around 6.7~keV, which appears consistently across all observations and can be naturally interpreted as an iron line from the reflection component. At the same time, noticeable variations in both the width and strength of this feature are observed. Within the framework of accretion disk reflection models, these variations suggest an evolution of the innermost accretion flow structure, which we will discuss in detail in Section~\ref{sec:discussion}.


Given that the source was in the disk-dominated state, we adopt a reflection model appropriate for thermal illumination, namely \texttt{reflionx\_bb} \footnote{https://github.com/honghui-liu/reflionx\_tables}, in our spectral fitting. For completeness, we also explore models including an additional corona-illuminated reflection component (\texttt{reflionx\_nth}). However, this extra component does not significantly improve the fit or alter the inferred spectral parameters; a detailed comparison is presented in Section~\ref{subsec:spectr_model}. \texttt{relflionx\_bb} is a rest-frame disk reflection model calculated with the reflionx code \citep{Ross2005}, which assumes a single-temperature blackbody as the incident radiation. It accounts for the reflection spectrum produced when thermal emission from the accretion disk re-illuminates its own surface as an effect of returning radiation \citep[e.g.][]{Dauser2022}. The incident blackbody temperature ($kT$) of \texttt{reflionx\_bb} are linked to $kT_{\rm in}$ of \texttt{diskbb}. Other free parameters of the reflection model include the ionization parameter and the iron abundance ($A_{\rm Fe}$). The convolution kernel \texttt{relconv} \citep{Dauser2010, Dauser2013} is required to include the relativistic effects. This model has parameters like the black hole spin ($a_{*}$), the disk inclination ($i$), and inner disk radius ($R_{\rm in}$). Previous studies have suggested a moderately to rapidly spinning black hole, with $a_{}\gtrsim0.7$ inferred from both continuum fitting \citep{Shafee2006} and reflection modelling \citep{Dong2020}, although these methods yield discrepant results \citep{Miller2009, Morningstar2014}. The broad iron line visible in Figure~\ref{nustar_residuals}--particularly in Epoch~1--also favours a high spin. Given this, and for the sake of reducing parameter degeneracy, we fix the spin at its maximal value ($a_{*}=0.998$) and proceed to constrain the inner disk radius $R_{\rm in}$.

\begin{figure*}
    \centering
    \includegraphics[width=0.98\linewidth]{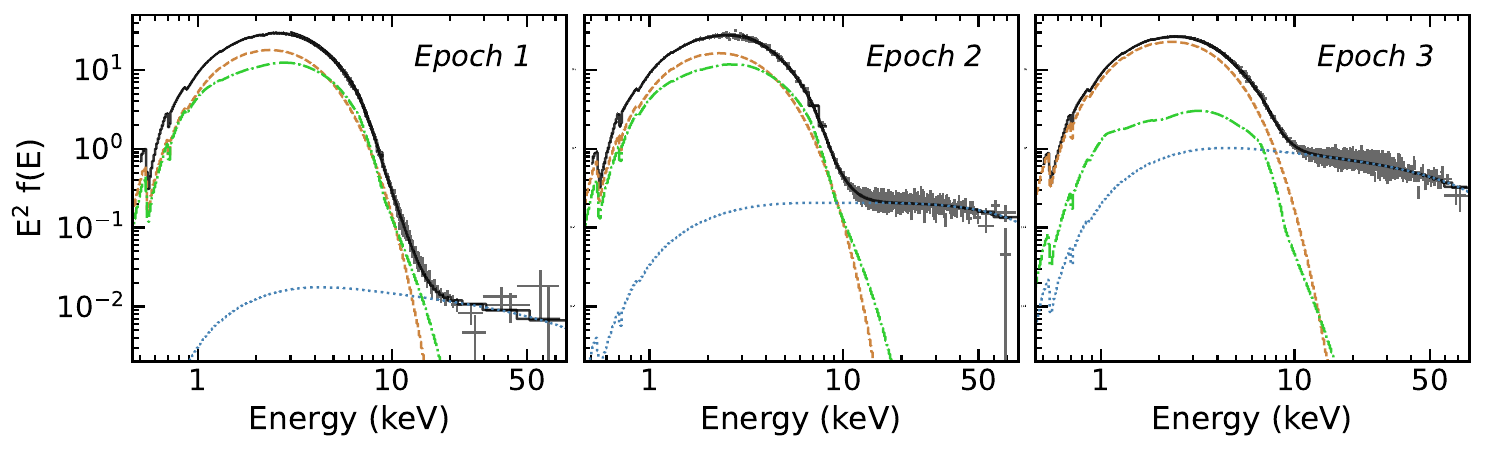} \\
    \vspace{0.1cm}
    \includegraphics[width=0.98\linewidth]{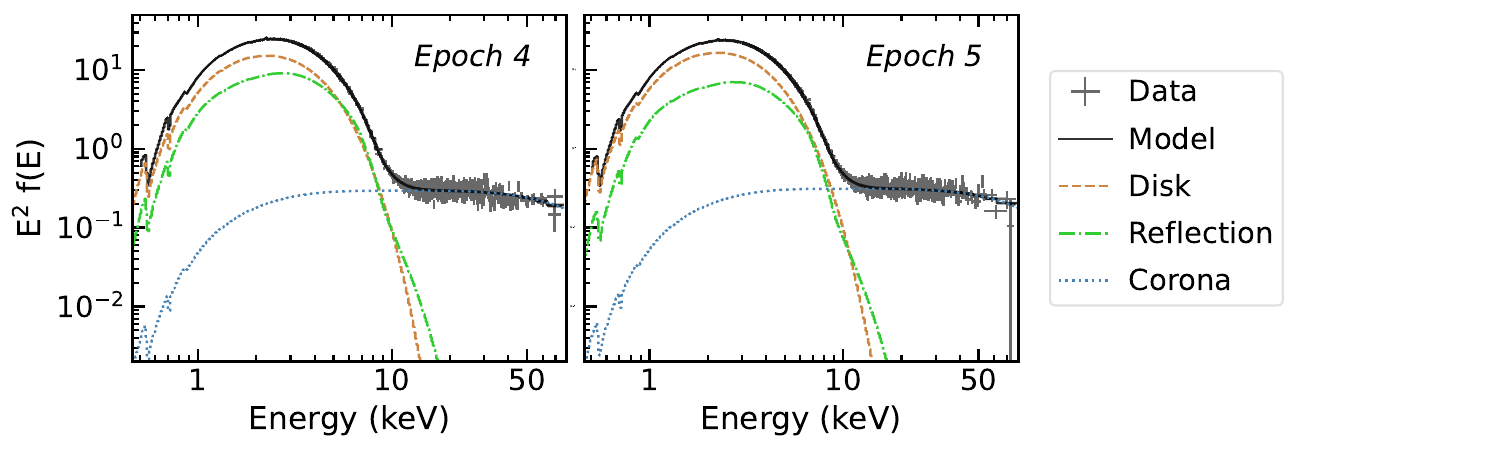} \\
    \caption{Broadband X-ray spectra and best-fitting models obtained from the five simultaneous \nustar\ and \nicer\ observations, corresponding to the vertical blue ticks in Figure~\ref{maxi_lc}. In each panel, the unfolded spectra [$E^2f(E)$; in units of keV$^2$~(photons~cm$^{-2}$~s$^{-1}$~keV$^{-1}$)] and the best-fitting model are displayed. The brown, green, and blue dashed curves represent the contributions from the disk, reflection, and Comptonized corona components, respectively.}
    \label{nustar_spectra}
\end{figure*}

We initially allow the inclination angle (\textit{i}) to vary freely during the fitting, and the resulting values are clustered between 15$^\circ$ and 30$^\circ$, which are consistent with previous study \citep{Zhang2025, Jin2024}. However, in some cases the parameter could not be well constrained, so we fix it at 20$^\circ$ in subsequent fits. A broken power-law emissivity profile is adopted. To better account for cross-calibration differences among instruments, we replace the simple \texttt{constant} model with the more flexible \texttt{Crabcorr} model \citep{Steiner2010} in the joint fitting. Noticeable residuals are present in the \nicer\ spectra at low energies, likely due to calibration issues \footnote{https://heasarc.gsfc.nasa.gov/docs/nicer/analysis\_threads/plot-ratio/} (e.g. \citealt{Zhang2024}). We therefore include a narrow emission line at 0.6~keV and a narrow absorption line at 2.2~keV to remove these residual features. This model combination successfully reproduces the broad-band spectra obtained from the simultaneous \nicer\ and \nustar\ observations, as shown in Figure~\ref{nustar_spectra}. The best-fitting parameters for Epochs~1–5 are listed in Table~\ref{nustar_fit}, and the posterior distributions of key parameters obtained for Epoch~3 are presented in Figure~\ref{epoch3_contour}. The hydrogen column density remains nearly constant around $N_{\rm H} \sim (3.8$–$4.2)\times10^{21}$~cm$^{-2}$, consistent with the Galactic value toward the source \citep{HI4PI-Collaboration2016}. The cross-calibration constants derived from the \texttt{Crabcorr} component remain within $\sim$5\%, confirming the good agreement between \nicer\ and \nustar\ flux calibration. 


Recent studies have shown that excess features in disk-dominated X-ray spectra of black hole X-ray binaries can be adequately described by including emission from within the plunging region, as traditional models tend to underestimate the flux in the intermediate energy range (6--10~keV) (e.g., \citealt{Fabian2020, Mummery2024}. Motivated by this, we test the inclusion of an additional blackbody component in the disk+corona model, but find that it does not provide an adequate fit. As illustrated in Figure~\ref{nustar_spectra}, the reflection component contributes excess emission at both low and high energies relative to the continuum, which cannot be accounted for by a single blackbody component.


\subsection{Spectral fitting results}

Given that the broadband spectra across all five epochs are well described by a combination of thermal disk emission, Comptonized continuum, and relativistic reflection, we attempt to extend the model combination to all available \hxmt\ and \nicer\ observations. Due to the narrower energy coverage and the relatively limited data quality of \hxmt\ and \nicer\ compared to \nustar, it is not feasible to constrain a large number of free parameters simultaneously. During the fitting of these spectra, we find that the emissivity profile was poorly constrained, and therefore we adopt a simple power-law profile for the emissivity. In cases where the Galactic hydrogen column density could not be constrained, owing to degeneracies with the thermal disk and reflection components, it is fixed at a value of $4\times10^{22}$~cm$^{-2}$. In general, this model combination provides an excellent description to \nicer\ and \hxmt\ spectra.

\begin{figure}
    \centering
    \includegraphics[width=1.0\linewidth]{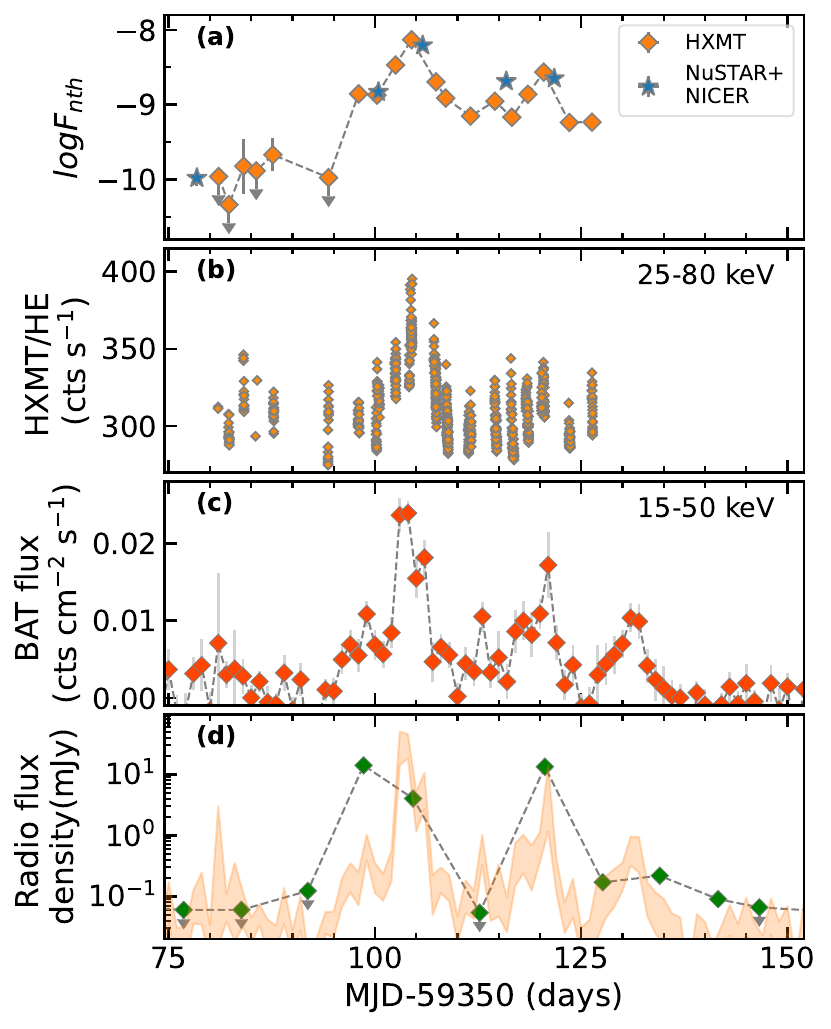} \\
    \caption{Evolution of the high-energy X-ray and radio fluxes. From top to bottom: Comptonized flux ($\log F_{\mathrm{nth}}$; erg~cm$^{-2}$~s$^{-1}$) derived from \hxmt\ (orange) and simultaneous \nustar+\nicer\ (blue) observations, the \hxmt/HE\ (25–80~keV) count rate, the \swift/BAT\ (15–50~keV) flux, and the 1.28~GHz radio flux density. The evolution of \swift/BAT\ flux is presented in panel (d) with the orange shaded area. }
    \label{nthcomp_flux}
\end{figure}

Figure~\ref{nthcomp_flux} illustrates the temporal evolution of the flux of Comptonized component, $F_{\mathrm{nth}}$, together with the high-energy X-ray and radio light curves during the radio flare events. Panel (a) shows the Comptonized flux ($\log F_{\mathrm{nth}}$) derived from \hxmt\ and simultaneous \nustar+\nicer\ fits. As a tracer of the coronal activity, $F_{\mathrm{nth}}$ exhibits pronounced variability, with its peak value reaching nearly two orders of magnitude higher than the initial level. The evolution of $F_{\mathrm{nth}}$ is consistent across different instruments. Panels (b) and (c) display the high-energy X-ray light curves from \hxmt/HE (25–80~keV) and Swift/BAT (15–50~keV), respectively, both tracing the hard X-ray emission from the corona. The \hxmt/HE spectra are not included in the spectral fitting; instead, we present the HE light curve to illustrate the variability of the emission intensity in this energy band. The consistent evolution of the HE and BAT light curves further supports the coronal flux trend shown in panel (a). Panel (d) shows the 1.28~GHz radio flux light curve, with the \swift/BAT light curve inserted as an indicator of the coronal activity. The radio flares coincide with the variations of the Comptonized flux, reinforcing the connection between the corona and jet.

\begin{figure}
    \centering 
    \includegraphics[width=1.0\linewidth]{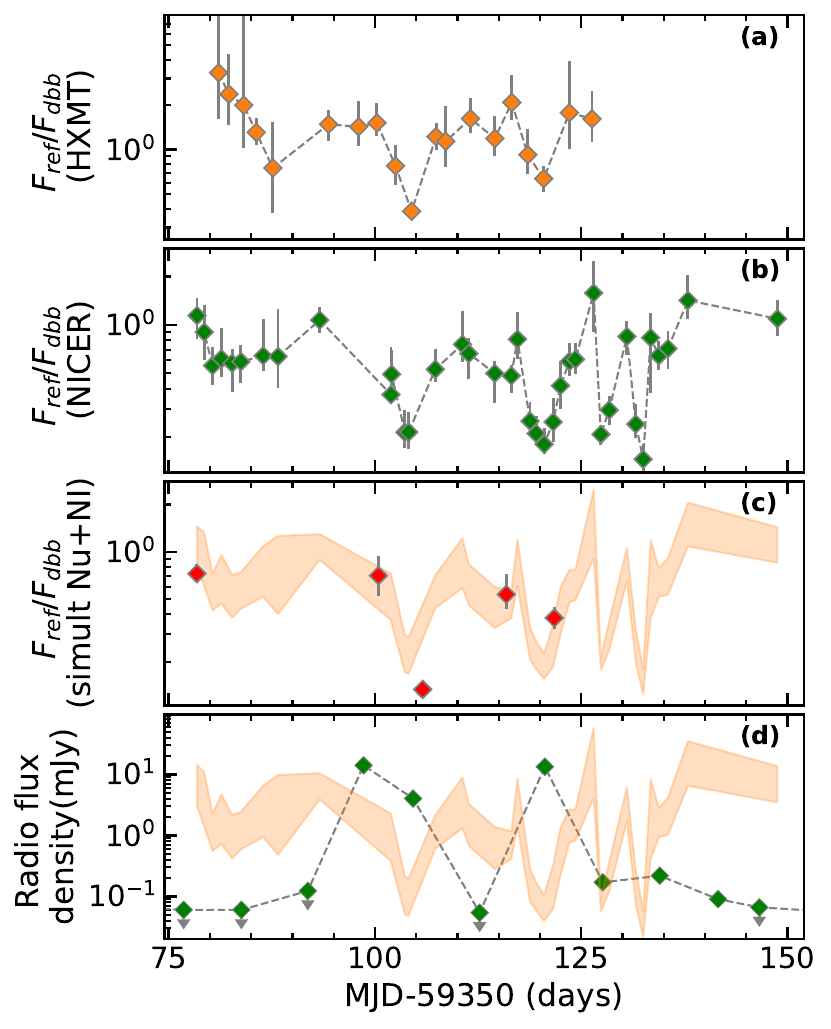} \\
    \caption{Evolution of the reflection-to-disk flux ratio ($F_{\mathrm{ref}}/F_{\mathrm{dbb}}$) and the radio flux density. From top to bottom: results from \hxmt, \nicer, and the simultaneous \nustar+\nicer\ fits, followed by the 1.28~GHz radio flux density. The temporal evolution of reflection fraction from \nicer\ spectra is inserted in panel (c) and (d) with the orange shaded area.}
    \label{refl_frac}
\end{figure}

Figure~\ref{refl_frac} shows the temporal evolution of the reflection-to-disk flux ratio ($F_{\mathrm{ref}}/F_{\mathrm{dbb}}$) derived from different instruments, together with the radio light curve. Panels (a)–(c) present the results obtained from \hxmt\ LE+ME, \nicer, and the simultaneous \nustar+\nicer\ observations, respectively. In panel (c), we insert the evolutionary trend extracted from  \nicer\ spectra. The two results exhibit a high level of consistency, indicating that the reflection fraction is robust across different datasets. In all cases, the reflection fraction ($F_{\mathrm{ref}}/F_{\mathrm{dbb}}$) displays significant variability over time, with its overall evolution broadly mirroring the radio light curve. In particular, panel (d) shows two dominant radio flares that coincide with pronounced dips in the reflection fraction, along with hints of additional weaker events. Within the framework of the returning radiation model, the fraction of thermal photons returning to the disk by relativistic effects is tightly correlated with the inner truncation radius of the accretion disk \citep{Dauser2022}. The ratio between the reflection and disk thermal fluxes ($F_{\mathrm{ref}}/F_{\mathrm{dbb}}$) could be a proxy of the inner disk radius. A lower reflection fraction implies a larger inner truncation radius of disk. Figure~\ref{refl_frac} shows that the X-ray reflection fraction undergoes significant temporal variations during the period of radio activity. While the reflection-fraction minima are not strictly coincident with the radio flares, and some occur after the flaring episodes, the overall evolution is suggestive of a connection between changes in the inner accretion flow geometry and the jet activity. The consistent behaviour observed with two independent X-ray instruments strengthens this interpretation.



\section{Discussion}\label{sec:discussion}

\subsection{Accretion state and Eddington ratio}

The radio flares detected by \meerkat\ occurred during a phase of steady decline in the X-ray flux. Assuming a distance of 7.5~kpc and a black hole mass of $9.4$~$M_{\bigodot}$, joint \nustar\ and \nicer\ fits show that the 0.01--100~keV X-ray luminosity decreased from $7.5\times10^{38}$~erg~s$^{-1}$ in Epoch~1 to $5.9\times10^{38}$~erg~s$^{-1}$ in Epoch~5, corresponding to a drop from about 60\% to about 50\% of the Eddington luminosity. \citet{Jin2024} performed a detailed analysis of all available \hxmt\ observations of the source and reported a transition from a geometrically thick accretion flow to a standard thin accretion disk occurring near the Eddington luminosity (MJD~$\sim 59400$). According to their study, the source was in a predominantly soft state during this period (MJD~$> 59428$), with the accretion disk well described by the standard thin-disk model, which is consistent with our spectral fitting results.

\subsection{Spectral modelling}\label{subsec:spectr_model}

In constructing our spectral model, we include three major components: the thermal emission from the accretion disk, the Comptonized emission from the corona, and the reflection component produced by a blackbody-illuminated disk. As shown in Figure~\ref{nustar_spectra}, the spectra are dominated by thermal emission, with only a relatively weak Comptonized coronal component detected across all datasets. This suggests that the reflection component is predominantly produced by returning radiation from the thermal emission of the accretion disk (see, e.g., \citealt{Mirzaev2024}). This interpretation is further supported by the fact that this model provides a good fit to the observed spectra. The strong correlation we observe between the hard X-ray power-law continuum and radio flux is reminiscent of the jet–corona connection proposed in models where hard X-rays originate in a magnetized jet base (e.g. \citealt{Markoff2004}). In such a scenario, if the primary X-ray source is beamed along the jet axis away from the disk, the observed reflection fraction would naturally be low. This could explain why the dominant reflection signature appears consistent with disk self-irradiation (returning radiation) rather than coronal illumination.

However, this configuration is not fully self-consistent, especially for observations with strong coronal emission, where the corona itself can contribute to the illuminating continuum and give rise to the reflection component. For instance, during Epoch~3 of the joint \nustar\ and \nicer\ observations, the Comptonized flux increased by nearly two orders of magnitude compared with Epoch~1, making it physically reasonable to consider disk reflection induced by the Comptonized emission.

To test this, we introduce an additional reflection component, \texttt{reflionx\_nth}, which adopts the \texttt{nthcomp} spectrum as its incident continuum, and apply the same relativistic convolution model (\texttt{relconv}) to account for relativistic effects. The parameters describing the relativistic blurring and disk properties are linked to those of the original blackbody-illuminated reflection component to ensure consistency. The fitting result shows that including this additional reflection component indeed improves the overall fit quality, $\Delta \chi^2 \sim 80$ with one extra free parameter. The improvement mainly occurs in the 0.5–2~keV and 20-79~keV energy ranges. However, the inclusion of this component does not significantly alter the constraints on most of the key physical parameters, and the primary impact is limited to parameters associated with coronal emission, with differences reaching up to a factor of $\sim$2, and to the measurement of the inner disk radius. The flux of this extra reflection component is only about one fifth—or even less—of the coronal emission in the high-energy band, and its contribution in the soft X-ray range is negligible compared to both the thermal disk emission and its reflected counterpart. When fitting the spectra of \nicer\ and \hxmt\ separately, it is difficult to constrain this component, so we do not include this component in the subsequent spectral modeling.

The variation of the inferred $R_{\rm in}$ caused by this additional reflection component demonstrates that this parameter is not uniquely constrained and remains dependent on the adopted spectral model. For this reason, we adopt the ratio of the reflection to disk fluxes ($F_{\mathrm{ref}}/F_{\mathrm{dbb}}$) as a more robust tracer of the inner truncation radius, rather than relying on the directly fitted values from spectra. We note that the measurement of the reflection fraction shows overall discrepancies between \hxmt\ and \nicer. This is mainly attributed to diﬀerences in the cross-calibration between instruments (see a more detailed discussion in \citealt{Zhao2024}).

Since \texttt{nthcomp} is an additive component, it may in principle bias the inferred disk flux and consequently affect reflection-related interpretations. To evaluate this potential effect, we test an alternative, more self-consistent treatment of Comptonization by replacing \texttt{nthcomp} with the convolution model \texttt{simplcutx} \citep{Steiner2017} applied to \texttt{diskbb}. We first apply this model to Epoch~3, where the coronal contribution is strongest. We find that this alternative model provides a comparable fit quality ($\chi^2 = 1619.63$ v.s. 1619.53 with \texttt{nthcomp}), with the overall spectral parameters remaining consistent with those obtained using \texttt{nthcomp}. The inferred disk flux shows only a minor increase, from $-7.073$ to $-7.049$ (in log units), corresponding to a difference of approximately 5\%. This indicates that while \texttt{nthcomp} may slightly underestimate the disk flux, the magnitude of this effect is small. The very low scattering fraction ($0.0529\pm0.003$) obtained with \texttt{simplcutx} further supports this conclusion. For the remaining observations, where the coronal component is weaker, the impact of this effect is expected to be even less significant. We therefore conclude that the use of \texttt{nthcomp} does not significantly affect the inferred spectral parameters and disk fluxes, nor does it alter any of the scientific conclusions presented in this work. Accordingly, we retain \texttt{nthcomp} in the spectral modeling throughout this work.

\subsection{Physical interpretation of disk truncation and jet launching}

The correlated evolution of the reflection fraction, Comptonized emission, and radio activity suggests a coupled instability involving the inner accretion disk, corona, and jet. In this framework, the observed decrease in the reflection fraction is interpreted as evidence for a transient outward retreat or truncation of the inner accretion disk, which could play a key role in enabling jet launching. A retreating inner disk may liberate magnetic energy through field line stretching and reconnection (e.g. \citealt{Hawley2015}), or alter the local flow geometry in a way that facilitates the formation of transient relativistic ejecta. At the same time, the partial evacuation of the inner disk region may enhance the energy supply to the corona—either via magnetic coupling between the disk and corona or through increased advection—providing a natural explanation for the contemporaneous rise in the Comptonized X-ray emission (e.g. \citealt{Nemmen2024}).

Within this picture, the radio flare is therefore linked to a broader disk–corona–jet coupling cycle, in which inner disk evacuation precedes or accompanies coronal brightening and episodic jet ejection. While the available data do not allow us to establish a strict causal sequence, the observed temporal associations are consistent with a scenario in which changes in the inner accretion flow geometry act as the common driver of both coronal and jet activity.

An alternative mechanism that could produce optically thin radio emission accompanied by harder X-ray emission is magnetic reconnection in the accretion disk (e.g., \citealt{Ripperda2020}). However, during the radio flaring episodes in this source, the peak flux density at 1.28 GHz reaches $\sim13-14$~mJy (and is likely even higher, as the weekly observing cadence may miss the true peak). Such a high flux implies a very large emitting region, with a characteristic physical size comparable to or exceeding that of the accretion disk itself \citep{Fender2006}. This makes a magnetic reconnection origin for the observed flares less favorable.

\subsection{Optical light curves}

\begin{figure}
    \centering 
    \includegraphics[width=1.0\linewidth]{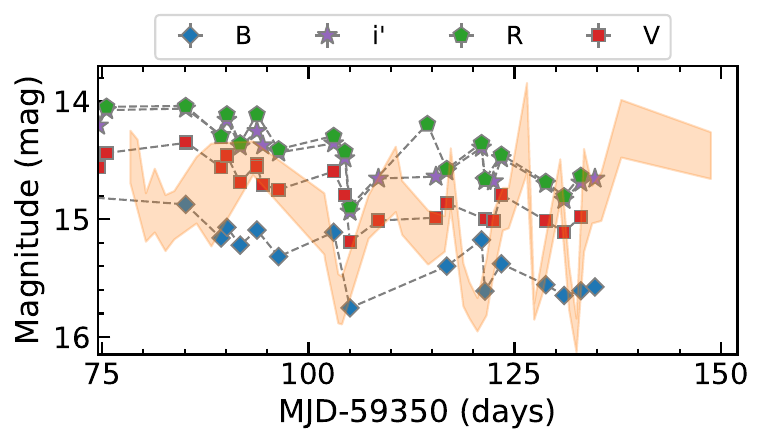} \\
    \caption{ The B- (blue), V- (red), R- (green), and i'-band (purple) light curves obtained with LCO. The shaded orange region represents the evolution of the reflection-to-disk flux ratio derived from \nicer\ spectra, which traces changes in the inner disk structure. }
    \label{opti_lc}
\end{figure}

During this outburst, intensive optical monitoring of the source was carried out with the 1-m and 2-m telescopes of the Las Cumbres Observatory (LCO), as a part of a monitoring campaign of $\sim50$ low-mass X-ray binaries since 2005--2007 \citep{Lewis2008}. All LCO photometric analysis and data calibration were performed using the "X-ray Binary New Early Warning System" pipeline (XB-NEWS; \citealt{Russell2019, Goodwin2020, Saikia2026}). The LCO optical data of \src\ are taken from Saikia et al. (in preparation), where the data reduction and calibration procedures will be described in detail. Figure~\ref{opti_lc} presents the optical light curves obtained in the i', R, V, and B bands, overlaid with the evolution of the reflection-to-disk flux ratio derived from the NICER spectra, which serves as an indicator of the inner disk truncation radius. The optical emission during outburst can arise from several possible mechanisms: X-ray reprocessing (e.g. \citealt{Vrtilek1990}), viscous heating in the accretion disk (e.g. \citealt{Du2025}), synchrotron emission from hot electrons in the inner flow (e.g. \citealt{Veledina2013}), and jet-related synchrotron radiation (e.g. \citealt{Tetarenko2015}). It has been proposed that optical emission primarily arises from the outer accretion disk due to X-ray reprocessing \citep{Russell2006}, where X-ray irradiation from the inner accretion flow illuminates and heats the outer accretion disk and produces optical emissions.

In Figure~\ref{opti_lc}, there are hints of a correlation between the optical fluxes and the inferred inner disk radius. Around MJD$\sim59455$, the reflection fraction shows a noticeable dip, which may indicate a transient increase in the inner disk radius, and this appears to coincide with a decrease in the optical fluxes. To quantify the correlation between the unevenly sampled inner disk radius and optical fluxes evolution curves, we computed the discrete correlation function (DCF; \citealt{Edelson1988}), which does not require interpolation and is well suited for irregularly sampled time series. The DCF results suggest a local maximum with a correlation coefficient of $\sim$0.5 near a lag of $\sim$1 day. Owing to the limited observational cadence, however, we cannot robustly establish whether any time delay exists between these variations, although the available data do not favour delays of several days. If real, such behaviour would be consistent with the optical emission containing a component produced by X-ray irradiation of the outer disk. The quasi-simultaneous decrease in optical flux would then point to possible structural changes in the accretion flow, although higher-cadence coverage would be required to confirm this interpretation. 




\subsection{X-ray timing properties}

\begin{figure}
    \centering 
    \includegraphics[width=1.0\linewidth]{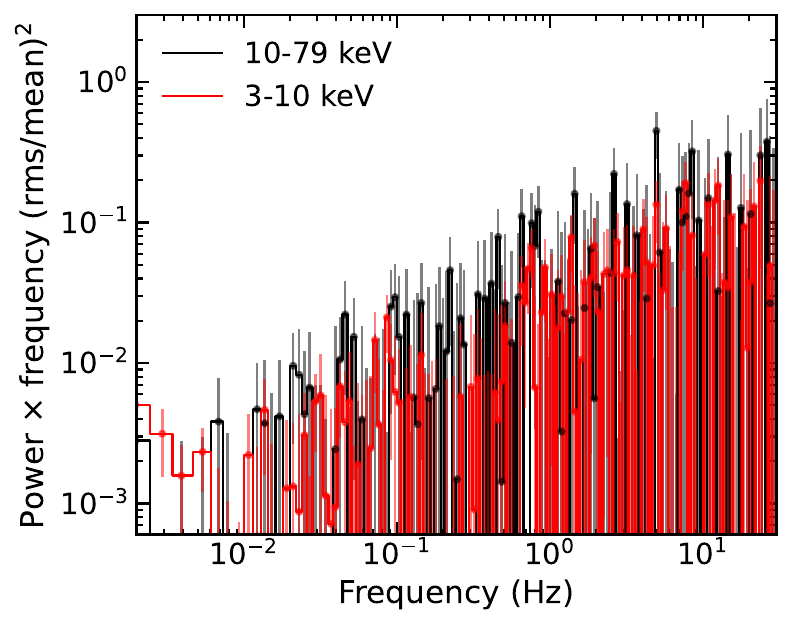} \\
    \caption{Energy-resolved cross power density spectra from the \nustar\ Epoch~3 observation, extracted in the 3–10~keV (disk-dominated; red) and 10–79~keV (corona-dominated; black) energy bands. Both spectra exhibit a flat, low-amplitude variability characteristic of the canonical disk-dominated state.}
    \label{epoch3_pds}
\end{figure}

To investigate possible changes in the X-ray timing properties during the radio flaring period, we examine all available \nicer\ epochs between MJD=59400 and MJD=59490. Across the entire set of observations, the power density spectra (PDS) consistently display low variability in the $5\times10^{-3}$--$50$~Hz frequency range, characteristic of the canonical soft state. We further generate energy-resolved PDS in the soft and hard X-ray bands with \nustar\ and \hxmt\ data to probe potential energy-dependent variability. For the \nustar\ data, we use the cospectrum derived from the two independent focal plane modules (FPMA and FPMB), defined as the real part of the cross power density spectra, as a proxy for the white-noise–subtracted PDS. This approach mitigates distortions of the white-noise level in the PDS caused by instrumental deadtime at high count rates \citep{Bachetti2015}. As an example, Figure~\ref{epoch3_pds} shows the cross PDS obtained from the \nustar\ Epoch~3 observation in the disk-dominated (3–10~keV) and corona-dominated (10–79~keV) bands. The two spectrum are nearly identical and both exhibit a featureless and low-variability shape. We have examined the PDS from the other \nustar\ epochs, together with the \hxmt\ and \nicer\ observations, and find that they display consistent overall shapes and timing characteristics across all datasets.

\begin{figure}
    \centering 
    \includegraphics[width=1.0\linewidth]{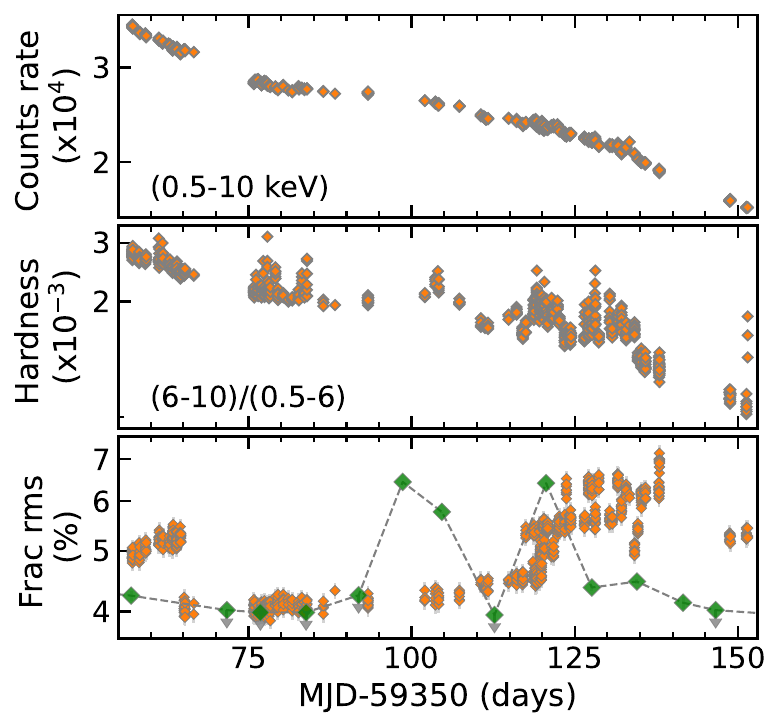} \\
    \caption{Evolution of the 0.5–10~keV \nicer\ count rate (top), spectral hardness defined as the (6–10~keV)/(0.5–6~keV) ratio (middle), and fractional rms variability (bottom) over the course of the episodic radio flares. The radio light curve obtained with \meerkat\ is inserted in bottom panel with green points. Each point represents a single \nicer\ snapshot, with grey vertical bars indicating the 1$\sigma$ uncertainties on the measurements. The source exhibits a smooth decay in X-ray flux accompanied by a gradual softening of the spectrum, superimposed with short-lived hardening excursions. The fractional rms remains relatively high during the early stages of the outburst, drops significantly as the source softens, and subsequently rises again at later times. Note that different \nicer\ observations during this interval were performed with different sets of active detectors. To avoid introducing additional uncertainties into the rms measurements due to these instrumental differences, we extracted only the events recorded by the subset of detectors that remained active throughout all relevant observations. The resulting light curves were then scaled to the equivalent count rate of 52 detectors before computing the rms.}
    \label{nicer_rms}
\end{figure}

We also check the evolution of \nicer\ count rate, hardness, and rms variability, which are shown in Figure~\ref{nicer_rms}. The rms is calculated over the frequency range 1/64--64~Hz. The Poisson noise level was estimated directly from the power spectrum by averaging the high-frequency powers above 64~Hz. The 0.5--10~keV count rate shows a smooth long-term decline, consistent with the overall decay of the outburst. Spectral hardness, defined as the ratio between the 6--10~keV and 0.5--6~keV bands, also decreases steadily as the source softens, exhibiting short-lived episodes of spectral variability superimposed on this trend. The fractional rms amplitude displays pronounced changes throughout the evolution. During the early portion (MJD$<$59415), the rms remains relatively high (5--6\%), before dropping to lower values as the source softens. In later stages (MJD$>$59470), the variability recovers and shows increased trend, indicative of enhanced short-timescale fluctuations. Although the drop in rms seems to occur before the onset of the radio flare, the two behaviours remain plausibly connected. Owing to the limited cadence and spatial resolution of the \meerkat\ observations, it is not possible to resolve the number of individual jet ejection events or determine their precise launch times; instead, we can only infer that jet activity appears to preferentially occur during periods of relatively low X-ray variability or correspond to a transition between low variability and high variability (a possible case for the radio flare around MJD$=$59470). A link between reductions in fast X-ray variability and the launch of discrete jet ejecta has been noted in several systems (e.g. \citealt{Fender2009, Miller-Jones2012, Carotenuto2025}). The source provides an illustrative example of this connection, supporting the notion of a broad coupling between the accretion flow and jet launching.



\subsection{Excursions toward a "harder" state}

The evolution of the source in the hardness–intensity diagram (HID) derived from the \nicer\ monitoring is shown in Figure~\ref{nicer_hid}. The HID displays a well-defined and monotonic track, with the 0.5--10~keV count rate strongly correlated with spectral hardness. As the outburst progresses, the source moves from the low-hardness, low-intensity region to progressively higher count rates and harder spectra, before turning over and softening again at later times. Two representative epochs, MJD~59426 and MJD~59499, are indicated in the diagram and mark the beginning and end of the \nicer\ observations used in our spectral analysis. The overall behaviour reflects a gradual transition toward softer X-ray spectra during the decline phase, superimposed with short-lived hardening excursions that appear broadly coincident with the times of the radio ejections, similar to the behaviour observed in XTE~J1859+226 \citep{Brocksopp2002}. These excursions are also evident in the MAXI HID diagram shown in Figure~\ref{maxi_hid}. The brief hardening excursions likely represent episodes where the inner disk becomes thermally/visciously unstable, leading to localized evaporation or truncation. This small-scale state change at the innermost region may be the direct precursor that channels energy into the corona and facilitates the discrete jet launch observed in the radio. However, since the temporal sampling of the \meerkat\ observations is relatively sparse, we cannot establish a strict one-to-one correspondence between the radio flares and the brief excursions of the source toward "harder" state in the HID.

\subsection{Comparison to jet ejection events revealed by \meerkat}

\begin{figure*}
    \centering 
    \includegraphics[width=1.0\linewidth]{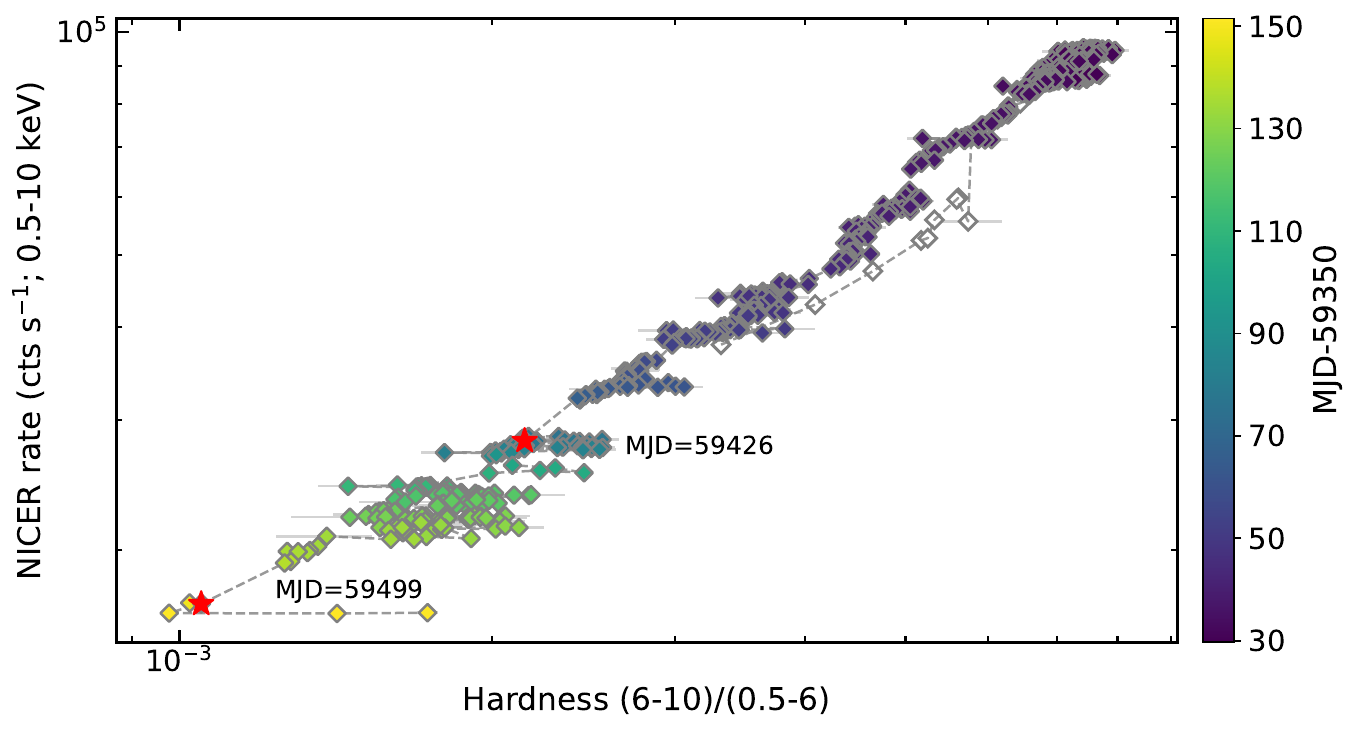} \\
    \caption{Hardness–intensity diagram of the source obtained from the \nicer\ observations. The hardness is defined as the ratio between the 6--10~keV and 0.5--6~keV count rates, while the intensity corresponds to the 0.5--10~keV count rate. Open (unfilled) diamonds correspond to observations during the rising phase of the outburst, whereas filled diamonds correspond to observations during the decay phase. Each point represents a single NICER snapshot. For the decay phase observation, the colour scale indicates the observation date in units of MJD$-59350$. Two representative points, MJD~59426 and MJD~59499, are highlighted to refer the beginning and end of the \nicer\ observations used in our spectral analysis.}
    \label{nicer_hid}
\end{figure*}

With \meerkat\ observations, \citet{Zhang2025_2} identified two distinct relativistic ejecta (E1 and E2) launched during the 2021 outburst of \src. The launch time of E1 is assumed to be MJD=59376, while that of E2 is inferred to be MJD=59503$\pm5$. Their launch times do not coincide with the times of the core radio flaring activity reported here. This suggests that the radio flares analyzed in our study probably correspond to $\geq 2$ jet ejection events with lower luminosity, which as a result are not identified later as spatially resolved by \meerkat, or that such discrete ejecta are present but become confused with E2 in the \meerkat\ epochs after MJD~$\sim59500$. The \nicer\ HID suggets that the source seemingly underwent one of its most pronounced short excursions toward "harder" state around MJD~$\sim59500$ (Figure~\ref{nicer_hid}), strikingly similar to that of the radio flares focused in this work. We do not explore it further, because the X-ray coverage around and after MJD~$\sim59500$ is limited and the corresponding radio flaring episode is not well monitored in X-rays band. We note that the inferred launch time of E2 is model-dependent. In particular, the earliest possible ejection time predicted by the model can be as early as MJD$\sim$59457, although it is not the statistically preferred case. This implies that we cannot fully exclude the possibility that the second radio flare corresponds to E2 reported in \citep{Zhang2025_2}.

\subsection{Towards an unified scenario of radio flares in disk dominated state of XRBs}

Similar radio flaring events during disk-dominated states have been reported in several X-ray binaries, such as MAXI~J1348--630 \citep{Carotenuto2025}, XTE~J1752--223 \citep{Brocksopp2013}, and EXO~1846--031 \citep{Williams2022}. Recently released ThunderKAT observations indicate that radio flares occurring during the soft state may be present in a larger number of sources \citep{Crook-Mansour2026}. In the case of MAXI~J1348--630, \citet{Carotenuto2025} proposed that short-lived compact jets were reactivated during an excursion to the hard-intermediate state and were switched oﬀ before the ejecta launch, and suggested a tentative correspondence between the launch of ejecta and the drop in X-ray rms variability. It is worth noting that, in MAXI~J1348--630, the decrease in fractional rms is accompanied by significant changes in the power density spectrum, including the appearance and disappearance of broadband noise components. In contrast, in \src, although a similar evolution in the fractional rms (i.e., a decrease followed by an increase) is observed, the shape of the power density spectrum does not show any significant changes. With the exception of MAXI~J1348--630, detailed studies of the simultaneous X-ray timing and spectral properties during these flaring episodes remain scarce in the literature. A systematic investigation of the X-ray timing and spectral behavior associated with such radio flares across different sources will be presented in a forthcoming work.


\section{Conclusions}\label{sec:conclusion}

We analyse the multi-wavelength properties of radio flares from \src\ during its disk-dominated state in 2021. With a comprehensive multi-wavelength study, we establish a direct observational link between the corona, inner accretion disk and jet activity in this system. The spectral analysis suggests that episodic radio flares occur contemporaneously with enhanced Comptonized X-ray emission and marked changes of the inner accretion flow. It demonstrates that the radio flares coincide with brief excursions toward "harder" state, together with a reduction in fractional rms variability. Our study implies a jet-launching mechanism that operates on short timescales within an otherwise soft accretion regime. Future coordinated multi-wavelength campaigns, with particularly high-cadence radio monitoring with AMI-LA or ATCA, will be essential for resolving the causal sequence linking disk truncation, coronal brightening, and jet launching in similar events.

\section*{Acknowledgements}

ZZ acknowledges support from ERC Synergy Grant 'Blackholistic' and the China Scholarship Council (CSC). RF acknowledges support from ERC Synergy Grant 'Blackholistic', The UKRI and The Hintze Family Charitable Foundation. This work makes use of observations from the LCO. DMR is supported by Tamkeen under the NYU Abu Dhabi Research Institute grant CASS. CB and JJ acknowledge support from the Warwick–Fudan Joint Seed Grant.

\section*{Data Availability}

All X-ray and radio data used in this study are publicly available. The \meerkat\ observations were performed as part of the ThunderKAT and X-KAT, which are large \meerkat\ open-time programmes to observe X-ray binaries in the radio band (PI: Rob Fender). The data can be accessed from \meerkat\ science archive (https://archive.sarao.ac.za/). The \nicer\ and \nustar\ observations can be accessed from the HEASARC science archive (https://heasarc.gsfc.nasa.gov/docs/archive.html), and the \hxmt\ data are available via the \hxmt\ science archive (http://HXMTweb.ihep.ac.cn).



\bibliographystyle{mnras}
\bibliography{example} 




\appendix

\section{Best-fit table and probability distribution obtained by MCMC}

Figure~\ref{nustar_spectra} shows that the broadband \nicer\ + \nustar\ spectra are well described by a model consisting of thermal disk emission, a Comptonized continuum, and relativistic reflection. The best-fitting parameters for Epochs~1–5 are listed in Table~\ref{nustar_fit} together with their 90\% confidence intervals derived from $\chi^2$ minimization.

\begin{figure*}
    \centering 
    \includegraphics[width=1.0\linewidth]{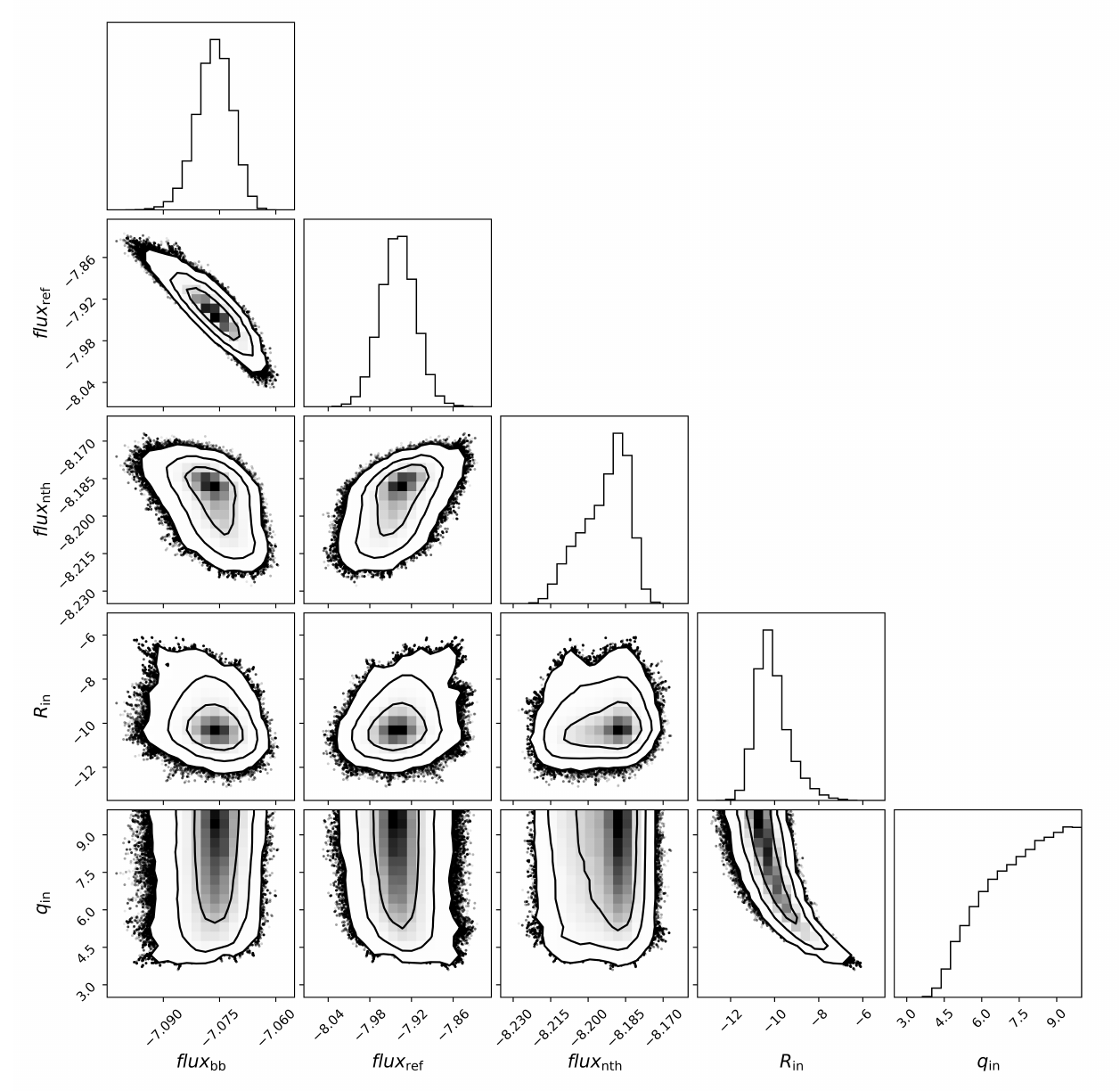} \\
    \caption{Correlations among the flux of the \texttt{diskbb} component, the flux of the reflection component, the flux of the corona component, the inner edge of the accretion disk $R_{\rm in}$ (in units of ISCO radius and with a minus sign, which is the output from \texttt{reflionx\_bb}), and the inner emissivity index $q_{\rm in}$ for Epoch~3. The three curves from the inside out represent, respectively, the 68\%, 95\%, and 99.7\% confidence level limits for two relevant parameters.}
    \label{epoch3_contour}
\end{figure*}

To further assess the robustness of the spectral constraints, we also estimated the parameter uncertainties using the Markov chain Monte Carlo (MCMC) method. We employed the Goodman–Weare algorithm with 100 walkers to generate a total of 40,000,000 samples, discarding the first 2,000,000 samples as burn-in to ensure convergence. As an example, the posterior distributions of key parameters obtained for Epoch~3 are presented in Figure~\ref{epoch3_contour}. Overall, the parameter uncertainties derived from the $\chi^2$ minimization are generally broader than those inferred from the MCMC analysis, indicating that the quoted confidence intervals are conservative.




\begin{table*}
\centering
\renewcommand\arraystretch{1.8}
\caption{Best-fit spectral parameters obtained from the joint \nicer\ and \nustar\ fits using the adopted reflection model. The model employed is \texttt{crabcorr $\times$ tbnew\_feo $\times$ (diskbb + relconv $\times$ reflionx\_bb + nthcomp + gauss) $\times$ gabs}, as described in Section~\ref{sec:analysis}. Uncertainties are quoted at the 90\% confidence level for a single parameter of interest. Parameters marked with an asterisk ($*$) are fixed during the fit. The symbols +P and $-$P denote that the upper or lower confidence bounds of the parameter are pegged at the limits of the allowed model parameter space.}
\begin{tabular}{lcccccc}
\hline\hline

Model \hspace{0.1cm} & \hspace{0.1cm} parameter \hspace{0.1cm} & \hspace{0.1cm} Epoch~1 \hspace{0.1cm} & \hspace{0.1cm} Epoch~2 \hspace{0.1cm} & \hspace{0.1cm} Epoch~3 \hspace{0.1cm} & \hspace{0.1cm}  Epoch~4 \hspace{0.1cm} & \hspace{0.1cm} Epoch~5 \hspace{0.1cm}  \\ \hline

\texttt{tbnew\_feo} & $N_{\rm H}$ [$10^{22}$~cm$^{-2}$] & $0.418_{-0.009}^{+0.010}$       & $0.387_{-0.011}^{+0.013}$      & $0.400_{-0.004}^{+0.004}$      & $0.380_{-0.004}^{+0.005}$      & $0.385_{-0.005}^{+0.005}$     \\
  & $A_{\rm O}$ & $0.96_{-0.03}^{+0.03}$         & $0.90_{-0.07}^{+0.07}$         & $0.77_{-0.04}^{+0.04}$         & $0.81_{-0.04}^{+0.05}$         & $0.78_{-0.03}^{+0.04}$        \\
  & $A_{\rm Fe}$ & ${1.0}^{*}$                    & ${1.0}^{*}$                    & ${1.0}^{*}$                    & ${1.0}^{*}$                    & ${1.0}^{*}$                   \\
\texttt{diskbb} & $T_{\rm in}$ [keV] & $0.8725_{-0.003}^{+0.0027}$    & $0.861_{-0.009}^{+0.008}$      & $0.861_{-0.004}^{+0.004}$      & $0.839_{-0.005}^{+0.005}$      & $0.840_{-0.004}^{+0.004}$     \\
  & $\log F_{\rm dbb}$ [erg~cm$^{-2}$~s$^{-1}$] & $-7.190_{-0.03}^{+0.024}$      & $-7.23_{-0.06}^{+0.06}$        & $-7.073_{-0.007}^{+0.007}$     & $-7.246_{-0.06}^{+0.029}$      & $-7.212_{-0.024}^{+0.022}$    \\
\texttt{relconv} & $q_{\rm in}$ & $4.35_{-0.15}^{+0.4}$          & $3.92_{-0.14}^{+1.3}$          & $10.0_{-5}^{+P}$               & $3.6_{-0.3}^{+0.3}$            & $10_{-6}^{+P}$               \\
  & $q_{\rm out}$ & $0.66_{-0.21}^{+2.9}$          & $0.4_{-P}^{+3.0}$              & ${1.0}^{*}$                    & $0.11_{-P}^{+2.3}$             & $3.05_{-0.29}^{+0.26}$        \\
  & $R_{\rm br}$ [R$_{\rm g}$] & $47_{-7}^{+20}$            & $73_{-68}^{+24}$             & ${80}^{*}$                   & $111_{-45}^{+18}$            & $6_{-P}^{+5}$               \\
  & $a_{\rm *}$ & ${0.998}^{*}$                    & ${0.998}^{*}$                    & ${0.998}^{*}$                    & ${0.998}^{*}$                    & ${0.998}^{*}$                   \\
  & $Incl$ [deg] & ${20.0}^{*}$                   & ${20.0}^{*}$                   & ${20.0}^{*}$                   & ${20.0}^{*}$                   & ${20.0}^{*}$                  \\
  & $R_{\rm in}$ [ISCO] & $2.48_{-0.16}^{+0.17}$        & $2.4_{-0.3}^{+0.3}$           & $10.8_{-2.8}^{+0.7}$          & $2.7_{-0.5}^{+0.7}$           & $3.6_{-0.5}^{+0.4}$          \\
\texttt{reflionx\_bb} & $\xi$ [erg~cm~s$^{-1}$] & $3431_{-458}^{+418}$           & $4301_{-1554}^{+751}$          & $2993_{-622}^{+605}$           & $4867_{-733}^{+673}$           & $5242_{-719}^{+618}$          \\
  & $A_{\rm Fe}$ & $10.0_{-1.3}^{+P}$             & $9_{-4}^{+P}$                & $10.0_{-0.7}^{+P}$             & $7.5_{-1.7}^{+1.2}$            & $8.8_{-1.4}^{+1.0}$           \\
  & $\log F_{\rm ref}$ [erg~cm$^{-2}$~s$^{-1}$] & $-7.329_{-0.024}^{+0.03}$      & $-7.38_{-0.07}^{+0.06}$        & $-7.96_{-0.04}^{+0.04}$        & $-7.52_{-0.07}^{+0.07}$        & $-7.63_{-0.05}^{+0.04}$       \\
\texttt{nthcomp} & $\Gamma$ & $2.2_{-0.3}^{+0.3}$            & $1.942_{-0.028}^{+0.03}$       & $2.22_{-0.03}^{+0.04}$         & $1.964_{-0.016}^{+0.021}$      & $1.948_{-0.020}^{+0.022}$      \\
  & $kT_{\rm e}$ [keV] & ${100.0}^{*}$                  & $30_{-P}^{+4}$               & $41_{-P}^{+57}$                & $30_{-P}^{+7}$               & $30_{-P}^{+8}$              \\
  & $\log F_{\rm nth}$ [erg~cm$^{-2}$~s$^{-1}$] & $-9.98_{-0.11}^{+0.15}$        & $-8.830_{-0.016}^{+0.014}$     & $-8.206_{-0.013}^{+0.015}$     & $-8.687_{-0.007}^{+0.009}$     & $-8.649_{-0.006}^{+0.009}$    \\
\texttt{gauss} & $E_{\rm line}$ [keV] & $0.646_{-0.010}^{+0.009}$       & $0.705_{-0.008}^{+0.007}$      & $0.721_{-0.008}^{+0.006}$      & $0.720_{-0.005}^{+0.005}$      & $0.724_{-0.007}^{+0.006}$     \\
  & $\sigma$ [keV] & $0.095_{-0.007}^{+P}$          & ${0.05}^{*}$                   & $0.050_{-0.011}^{+0.013}$      & $0.043_{-0.013}^{+0.014}$      & $0.041_{-0.013}^{+0.014}$     \\
  & Norm [photons~cm$^{-2}$~s$^{-1}$] & $3.3_{-0.4}^{+0.5}$            & $0.98_{-0.22}^{+0.25}$         & $0.75_{-0.14}^{+0.22}$         & $0.58_{-0.11}^{+0.15}$         & $0.53_{-0.1}^{+0.16}$         \\
\texttt{gabs} & $E_{\rm line}$ [keV] & $2.21_{-0.03}^{+0.03}$         & ${2.2}^{*}$                    & ${2.2}^{*}$                    & $2.228_{-0.022}^{+0.022}$      & $2.25_{-0.07}^{+P}$           \\
  & $\sigma$ [keV] & $0.05_{-0.05}^{+0.05}$         & ${0.01}^{*}$                   & ${0.01}^{*}$                 & ${0.01}^{*}$          & $0.021_{-0.019}^{+P}$         \\
  & Norm [keV] & $0.0041_{-0.0017}^{+0.0019}$   & ${0.0}^{*}$                    & ${0.0}^{*}$                    & $-0.0010_{-0.0026}^{+0.0006}$  & $-0.0016_{-0.0019}^{+0.0014}$ \\
\texttt{crabcorr} & $\Delta \Gamma_{\rm FPMA}$ & $0.042_{-0.007}^{+0.007}$      & $0.071_{-0.020}^{+0.020}$         & $0.040_{-0.009}^{+0.009}$      & $0.032_{-0.007}^{+0.010}$       & $0.041_{-0.009}^{+0.009}$     \\
  & Norm & $1.106_{-0.011}^{+0.011}$      & $1.120_{-0.026}^{+0.027}$      & $1.060_{-0.012}^{+0.013}$      & $1.037_{-0.010}^{+0.014}$       & $1.051_{-0.013}^{+0.013}$     \\
\texttt{crabcorr} & $\Delta \Gamma_{\rm FPMB}$ & $0.041_{-0.007}^{+0.007}$      & $0.052_{-0.020}^{+0.020}$         & $0.020_{-0.009}^{+0.009}$      & $0.012_{-0.007}^{+0.005}$      & $0.037_{-0.009}^{+0.009}$     \\
  & Norm & $1.061_{-0.011}^{+0.011}$      & $1.065_{-0.025}^{+0.026}$      & $1.007_{-0.012}^{+0.012}$      & $0.990_{-0.010}^{+0.007}$       & $1.021_{-0.013}^{+0.013}$     \\
  \hline
  & $\chi^2/dof$ & $1081.65/1298$ & $1440.99/1449$ & $1619.53/1791$ & $1481.02/1640$ & $1279.20/1662$  \\
\hline\hline
\end{tabular}
\label{nustar_fit}
\end{table*}



\section{MAXI hardness-intensity diagram}

\begin{figure}
    \centering 
    \includegraphics[width=1.0\linewidth]{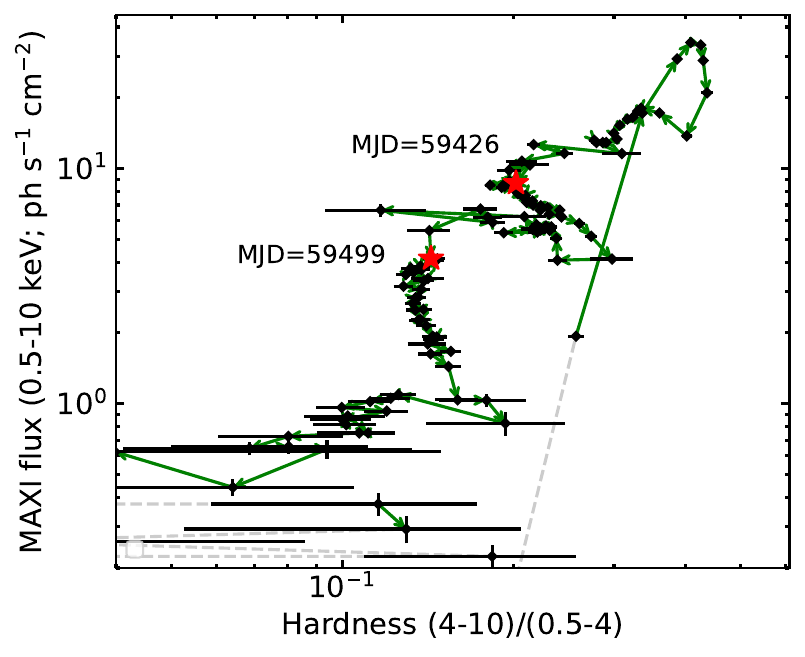} \\
    \caption{Hardness-intensity diagram of \src\ with MAXI daily average data. Two representative points, MJD 59426 and MJD 59499, are highlighted as in Figure~\ref{nicer_hid}.}
    \label{maxi_hid}
\end{figure}

Figure~\ref{maxi_hid} presents the hardness–intensity diagram constructed from the MAXI observations. The data points corresponding to MJD 59426 and 59499 are highlighted with red stars. During this interval, the source underwent brief excursions toward the hard state before returning to the soft state.


\bsp	
\label{lastpage}
\end{document}